\documentclass[aps,pra,twocolumn,showpacs,longbibliography,superscriptaddress]{revtex4-1}  % for review and submission
\usepackage{graphicx}  % needed for figures
\usepackage{dcolumn}   % needed for some tables
\usepackage{bm}        % for math
\usepackage{amsmath}
\usepackage{amsthm}
\usepackage{amssymb}
\usepackage{bbold} % to generate identity 1
\usepackage{mathrsfs}
\usepackage{array}
\usepackage[dvipsnames,usenames]{color}
\usepackage{soul} % to use \st for strikeout 
\usepackage{ulem} % to use \sout for strikeout to include equations
\usepackage{physics}
\usepackage{ulem}

\usepackage{color}
\usepackage[dvipsnames,usenames]{color}

\usepackage{braket}
\usepackage{float}
\usepackage{mhchem}
\usepackage[makeroom]{cancel}
\usepackage{upgreek}
\usepackage{commath}
\usepackage{siunitx}
\usepackage{orcidlink}

\usepackage[titletoc]{appendix}

\usepackage{hyperref}
\hypersetup{colorlinks=true, linkcolor=magenta, citecolor=magenta, urlcolor=magenta}

\begin{document}
\normalem

\title {The Two-Dimensional Rashba-Holstein Model: A Quantum Monte Carlo Approach}

\author{Julián Faúndez \orcidlink{0000-0002-6909-0417}}
\affiliation{Instituto de F\'isica, Universidade Federal do Rio de Janeiro Cx.P. 68.528, 21941-972 Rio de Janeiro RJ, Brazil}
\affiliation{Departamento de Física y Astronomía, Universidad Andres Bello, Santiago 837-0136, Chile}
\author{Rodrigo A. Fontenele \orcidlink{0000-0002-8251-0527}}
\affiliation{Instituto de F\'isica, Universidade Federal do Rio de Janeiro Cx.P. 68.528, 21941-972 Rio de Janeiro RJ, Brazil}
\author{S. dos A. Sousa-Júnior \orcidlink{0000-0002-4266-3780}}
\affiliation{Department of Physics, University of Houston, 77204, Houston, Texas, USA}
\author{Fakher F. Assaad \orcidlink{0000-0002-3302-9243}}
\affiliation{Institut f\"ur Theoretische Physik und Astrophysik, Universit\"at W\"urzburg, 97074 W\"urzburg, Germany}
\affiliation{W\"urzburg-Dresden Cluster of Excellence ct.qmat, Am Hubland, 97074 W\"urzburg, Germany}
\author{Natanael C. Costa \orcidlink{0000-0003-4285-4672}}
\affiliation{Instituto de F\'isica, Universidade Federal do Rio de Janeiro Cx.P. 68.528, 21941-972 Rio de Janeiro RJ, Brazil}
\affiliation{Institut f\"ur Theoretische Physik und Astrophysik, Universit\"at W\"urzburg, 97074 W\"urzburg, Germany}

%\date{\today}

\begin{abstract}
In this work, we investigate the impact of Rashba spin-orbit coupling (RSOC) on the formation of charge-density wave (CDW) and superconducting (SC) phases in the Holstein model on a half-filled square lattice. Using unbiased finite-temperature Quantum Monte Carlo simulations, we go beyond mean-field approaches to determine the ground state order parameter as a function of RSOC and phonon frequency. Our results reveal that the Rashba metal is unstable due to particle-hole instabilities, favoring the emergence of a CDW phase for any RSOC value. In the limit of a pure Rashba hopping, the model exhibits a distinct behavior with the appearance of four Weyl cones at half-filling, where quantum phase transitions are expected to occur at strong interactions. Indeed, a quantum phase transition, belonging to the  Gross-Neveu Ising universality class between a semi-metal and CDW   emerges at finite phonon frequency dependent coupling $\lambda_c$. In the antiadiabatic limit we observe an enhanced symmetry in the infrared that unifies SC and CDW orders.  These results advance our understanding of competing CDW and SC phases in systems with spin-orbit coupling, providing insights that may help clarify the behavior of related materials.
\end{abstract}


\maketitle

\section{Introduction}

The interplay between spin-orbit coupling (SOC) and electron-phonon coupling (EPC) remains an open issue in condensed matter physics. Even for the single-excitation case, the leading effects of the competition between the energy gained from coupling with phonons (and its quasiparticle formation) and the electronic dispersion from the SOC are subtle, with the effective polaron+ mass depending on both EPC and SOC~\cite{Cappelluti2007,Covaci2009,Li2010,Li2011}. This delicate balance could have direct implications for the development of novel devices, particularly in the field of spintronics\,\cite{Zutic2004}.
The degree of complexity is quite increased in the many-electron case, due to the inclusion of electronic correlations. As strong electron-electron interactions introduce many-body effects that may lead to long-range ordered phases, the SOC changes the Hamiltonian's symmetries and, consequently, the electrons' dynamics. This fundamental change could lead to unpredictable physical responses or new phases of matter\,\cite{Pesin10}. Some iridates or pyrochlores materials are prototypes for such interplay, as they are Mott insulators and also exhibit strong SOC effects. Indeed, these compounds show a myriad of phases, from Weyl semimetals and topological insulators, to spin liquid phases\,\cite{Krempa14,Rau16}. However, while the interplay between SOC and direct (Coulomb) electron-electron has been vastly scrutinized in the literature\,\cite{Riera2013,Laubach2014,Sun2017,Kennedy2022,Kawano2023,Beyer2023,Gastiasoro2023,Beyer2023}, the leading effects of SOC on indirect interactions, such as the EPC, and vice versa, are much less clear.

This interplay can be explored in some transition-metal dichalcogenides (TMDs), such as 2H-TaSe$_{2}$ and 2H-TaS$_{2}$, which exhibit significant EPC and SOC effects\,\cite{Liu16, Manzeli17, Zhang24}. In the former, both in bulk and single-layer forms, the charge-density wave (CDW) phase in the ground state appears to be barely affected by SOC, with the wavevector of this phase remaining almost unchanged, whether SOC is present or not\,\cite{Ge2012}. In contrast, in 2H-TaS$_{2}$, SOC has a pronounced impact on the EPC strength, suppressing EPC matrix elements. This suppression of EPC by SOC is essential for accurately explaining the superconducting critical temperatures observed in this material\,\cite{Lian2022}.
Another interesting example is the Pb crystal, which is a conventional (strong coupling) superconductor. For this material, the EPC strength, calculated from Eliashberg functions, is drastically affected by the inclusion of SOC, which is essential to replicate the experimentally observed superconducting (SC) properties\,\cite{Heid2010}. Finally, the interplay between charge order and SOC remains a subject of ongoing debate even for 1D systems, such as in atomic wire arrays on semiconductor surfaces\,\cite{Snijders2010}, or in the quasi-1D material (TaSe$_4$)$_{2}$I. The latter, in particular, is considered as a Weyl semimetal\,\cite{Shi2021}, exhibiting Rashba-like band splittings, but also undergoes a CDW phase transition at $T_{\rm CDW}=263$ K\,\cite{Zhang2020}.

\begin{figure*}[t]
    \centering
    \includegraphics[scale=0.35]{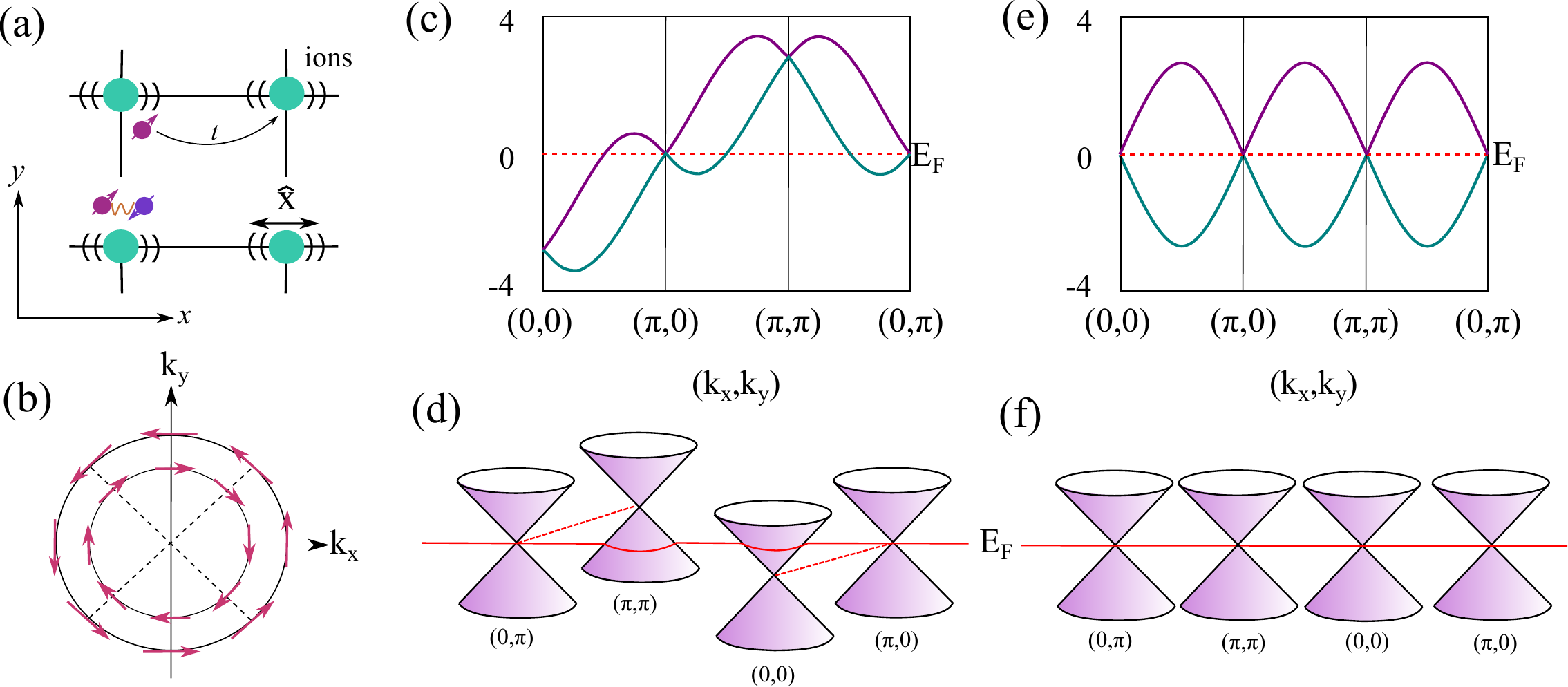}
    \caption{(a) Sketch of the Holstein model: vibrations of lattice ions, represented by local dispersionless phonons, mediate an indirect electron-electron interaction. (b) Illustration of spin splitting due to Rashba spin-orbit coupling in a 2D electron gas, producing spin textures with opposite chiralities. (c) The electronic dispersion relation along the high-symmetry points in a square lattice with finite Rashba spin-orbit coupling. (d) Pictorial representation of four Weyl cones, with two located at the Fermi level (indicated by the red line). (e) Same as in panel (c), but in the pure Rashba hopping limit where $t/\alpha=0$. (f) Same as in panel (d), but in the pure Rashba hopping limit where $t/\alpha=0$.}
    \label{Fig1}
\end{figure*}

From a theoretical perspective, the Holstein model captures the essential features of systems where EPC plays a central role\,\cite{Holstein1959}. This model describes local (dispersionless) harmonic oscillators at each site of the lattice, which couple locally to the electronic density, as illustrated in Fig.\,\ref{Fig1}\,(a). 
The Holstein model has been extensively studied in the literature, exhibiting a competition between CDW and SC phases that depends on finely tuned external parameters or the lattice geometry\,\cite{Li2015,Costa2018,Dee2019,Zhang20192,Chen2019,Cohen-Stead2019,Dee2020,Bradley2021,Xiao2021,Costa2021,Araujo2022,Weber2022,SASJR2023}. However, to the best of our knowledge, the many-body problem of interacting electron-phonon coupled electrons in the presence of SOC has only recently been explored, particularly within the context of the Holstein Hamiltonian.
Indeed, Ref.\,\onlinecite{Fontenele2024} investigates the impact of Rashba spin-orbit coupling (RSOC) on the stability of the CDW phase within a mean-field approximation, while Ref.\,\onlinecite{Xiangyu2024} employs an unbiased Quantum Monte Carlo (QMC) approach to explore the emergence of an induced spin-orbit coupling (SOC) when the EPC is spin-dependent, both in the half-filled square lattice. The antiadiabatic limit has also been analyzed in Refs.\,\onlinecite{Tang2014,Rosenberg2017,Song2024} using QMC methods, as further discussed below.

Motivated by these experimental and theoretical results, we further investigate how a RSOC affects a system with strong EPC, analyzing it both for the adiabatic and antiadiabatic limits. To this end, we employ unbiased finite-temperature QMC methods \cite{Blankenbecler1981,Hirsch1983,Hirsch1985,Scalettar1989}, examining the charge and pairing correlations of the two-dimensional Rashba-Holstein model at the half-filling of the square lattice. As our main result, we provide finite temperature and ground state phase diagrams. This article is organized as follows: in the next section we present the model and its symmetry properties, as well as the methodology; in section \ref{results}, we present and discuss the our results, leaving our conclusions and further remarks to section \ref{conclusion}.

\color{black}

\section{Model and methodology}\label{model}

\subsection{The model}

In this work, we investigate the properties of the Rashba-Holstein model (RHM), whose Hamiltonian reads
\begin{align} 
    \mathcal{H} &= -t\sum_{\langle \mathbf{i},\mathbf{j} \rangle, \sigma}  \left( c_{\mathbf{i},\sigma}^{\dagger} c_{\mathbf{j},\sigma} + {\rm H.c.} \right) - \mu \sum_{\mathbf{i},\sigma} n_{\mathbf{i},\sigma} \nonumber \\
    & - \alpha \sum_{\mathbf{i},\sigma,\sigma^{\prime}} \bigg[ \big( i c_{\mathbf{i},\sigma}^{\dagger} [\hat{\sigma}^{x}]_{\sigma, \sigma^{\prime}} c_{\mathbf{i}+\mathbf{\hat{y}},\sigma^{\prime}} - i c_{\mathbf{i},\sigma}^{\dagger} [\hat{\sigma}^{y}]_{\sigma, \sigma^{\prime}} c_{\mathbf{i}+\mathbf{\hat{x}},\sigma^{\prime}}\big)  \nonumber\\
    & + {\rm H.c.}\bigg] + \sum_{\mathbf{i}} \left( \frac{\hat{P}_{\mathbf{i}}^{2}}{2M} + \frac{M \omega_{0}^{2}}{2} \hat{X}^{2}_{\mathbf{i}} \right) - g \sum_{\mathbf{i},\sigma} \big(n_{\mathbf{i},\sigma} - 1\big) \hat{X}_\mathbf{i} ,
    \label{Eq1}
\end{align}
where sums run over a two-dimensional square lattice under periodic boundary conditions, with \(\langle \mathbf{i},\mathbf{j} \rangle\) denoting nearest neighbors sites. Here, we use the standard second quantization formalism, with \( c_{\mathbf{i},\sigma}^{\dagger} \) (\( c_{\mathbf{i},\sigma} \)) being creation (annihilation) operators of electrons in a given site $\mathbf{i}$ with spin \(\sigma (= \uparrow, \downarrow)\), while $n_{\mathbf{i},\sigma} \equiv c_{\mathbf{i},\sigma}^{\dagger} c_{\mathbf{i},\sigma}$ describe the number operators. The first term on the right hand side of Eq.\,\eqref{Eq1} corresponds to the non spin-flip contribution to the kinetic energy of electrons, with \( t \) being the hopping integral, while the second term is the chemical potential $\mu$. The third term describes the spin-orbit contribution, of strength $\alpha$, with $[\hat{\sigma}^{\gamma}]_{\sigma, \sigma^{\prime}}$ defining the elements of Pauli matrices ($\gamma=x$, $y$ and $z$). The fourth term denotes the local harmonic oscillators (HO) of mass $M$ and frequency $\omega_{0}$, with \(\hat{X}_{\mathbf{i}}\) and \(\hat{P}_{\mathbf{i}}\) being the conjugate displacement and momentum operators. The last term in Eq.\,\eqref{Eq1} describes the on-site electron-phonon coupling of strength $g$. Hereafter, we define the lattice constant and the HO's masses as unities, while setting the energy scale by the hopping integral, $t$.

Before proceeding, it is worthy discussing the limit cases. For instance, the occurrence of long-range order in the pure Holstein model ($\alpha =0$) has been under intense investigation over the past years, at different lattices and fillings. In particular, in the half-filled square lattice, there are strong evidences that support the emergence of CDW for any EPC $g > 0$, at any frequency $\omega_{0}$\,\cite{Weber17a,Hohenadler19,Costa2020}, different from the 1D case, which exhibits a quantum phase transition\,\cite{Jeckelmann99,Hohenadler2018}.
In the antiadiabatic limit, the phonon degree of freedom can be integrated out\,\cite{Berger95}, and the Holstein model is mapped onto the attractive Hubbard model, with effective dynamic electron-electron interaction being $U_{\rm eff} (\omega) =  - \frac{g^{2} / \omega^{2}_{0}}{1 - (\omega /\omega_{0})^2 }$. We define $g^{2} / \omega^{2}_{0} \equiv \lambda$, which is the energy scale for the polaron formation, and sets the effective attractive interaction strength in this limit.
Notice that the mapping into the attractive Hubbard model leads to a coexistence between CDW and SC for any $|U|>0$, enforced the $SU(2)$ symmetry.

We also recall that the mean-field analysis of the Holstein model with RSOC, discussed in Ref.\,\onlinecite{Fontenele2024}, (or any electron-phonon model) corresponds to the limit in which the lattice is already distorted. That is, when the kinetic term of the quantum HO's are neglected. Though interesting, this limit it is essentially classical, as the Hamiltonian does not contain non-commuting terms. Therefore, the mean-field results are only relevant when dealing with low frequencies, but only qualitatively. Treating quantum fluctuations by adding momentum operators is nontrivial,, and the resulting behavior may be unexpected. One could approach it perturbatively using the Wigner-Kirkwood method (or truncated Wigner approximation)\,\cite{Steel98,Polkovnikov03,Paprotzki24}, but this requires keeping $\hbar \omega_0 / k_{\rm B}T$ small. In our work, by contrast, we employ a nonperturbative (unbiased) methodology, which constitutes a significant improvement over any mean-field or perturbative approach.
As mentioned in the Introduction, there are only a few recent unbiased studies available on two-dimensional strongly interacting systems in the presence of SOC. Reference \onlinecite{Tang2014} showed that the critical temperatures of the attractive Hubbard model is drastically enhanced by the inclusion of a RSOC, while Refs.\,\onlinecite{Rosenberg2017,Song2024} investigate how the symmetry of Cooper pair is affected by the RSOC. Both references analyzed the antiadiabatic limit (i.e.~for the attractive Hubbard model), and away from the WSM case.

\subsection{Symmetries}
\label{SubSection:Symmetries}

We first note that the Rashba term breaks inversion (parity) symmetry. The inversion operation is represented by a unitary operator $\Pi$ with $\Pi\, \mathbf{r}\, \Pi^{-1} = -\mathbf{r}$; in momentum space, $\Pi\, \mathbf{k}\, \Pi^{-1} = -\mathbf{k}$. The spin operators are left unchanged by $\Pi$. Given that the Rashba spin-orbit term in the square lattice is given by
$$
H_R(\mathbf{k}) = 2\alpha \bigl(\sin k_y\, \sigma_x - \sin k_x\, \sigma_y \bigr),
$$
it is clear that
$$
\Pi\, H_R(\mathbf{k})\, \Pi^{-1} = H_R(-\mathbf{k}) = - H_R(\mathbf{k}),
$$
showing that $H_R$ is odd in $\mathbf{k}$ and not invariant under $\Pi$. Thus, the full Hamiltonian lacks inversion symmetry for any $\alpha \neq 0$.

In absence of EPC, the most relevant effect of the inclusion of a RSOC is the spin-dependent momentum shift, which splits the spin textures, with well-defined chiralities, as illustrated in Fig.\,\ref{Fig1}\,(b). It results in a (chiral) splitting of the band structure, shifting the van Hove singularity away from the half-filling, as depicted in Fig.\,\ref{Fig1}\,(c).
Where the split branches cross, the spectrum exhibits a Weyl cone, which acts as a source or sink of Berry curvature (nonzero net flux through a small enclosing surface). For the square lattice, four Weyl cones emerge: two at the Fermi level [$(\pi,0)$ and $(0,\pi)$], and two others above and below the Fermi level [at $(\pi,\pi)$ and $(0,0)$, respectively], as illustrated in Fig.\,\ref{Fig1}\,(d).
The position of the Weyl cones may change depending on the ratio $\alpha /t$, with all cones being at the Fermi level in the limit case of $t/\alpha = 0$ (pure Rashba), where a Weyl semimetal (WSM) emerges; see, e.g., Figs.\,\ref{Fig1}\,(e) and (f).
As discussed below, the spin-dependent momentum shift due to the RSOC preserves the Fermi surface nesting, creating particle-hole instabilities for any finite ratio $t/\alpha$\,\cite{Kawano2023,Kubo2024}.

At vanishing chemical potential, $\mu = 0$, the Hamiltonian is invariant under combined particle-hole symmetry
\begin{equation}
\mathcal P^{-1}\, c_{\mathbf{i},\sigma}^\dagger\, \mathcal P = \eta_{\mathbf{i}} c_{\mathbf{i},\sigma},
\end{equation}
and canonical transformation 
\begin{equation}
    \hat{X}_{\mathbf{i}} \rightarrow - \hat{X}_{\mathbf{i}}, \quad 
    \hat{P}_{\mathbf{i}} \rightarrow - \hat{P}_{\mathbf{i}},
\end{equation}
with $\eta_{\mathbf{i}} = e^{i \mathbf{i}\cdot\mathbf{Q}}$, in which $\mathbf{Q} = (\pi,\pi)/a $ corresponds to the antiferromagnetic wavevector. 
As a consequence, $\mu=0$ corresponds to  half-band filling, $\langle n_{\mathbf{i}}\rangle  =1$.   In the absence of  interactions,  the energy bands read: 
\begin{equation}
    E_{\pm}(\mathbf{k})  =   \epsilon(\mathbf{k}) \pm | \mathbf{V}(\mathbf{k}) | 
\end{equation}
with $ \epsilon(\mathbf{k}) = -2 t \left[ \cos(k_x) +    \cos(k_y)  \right] $, and 
$\mathbf{V}(\mathbf{k}) = 2 \alpha \left[ - \sin(k_y), 
\sin(k_x),0\right]$.  For the antiferromagnetic wave  $\mathbf{Q}$  and at half-filling, $\mu = 0$, the nesting condition  is satisfied as
\begin{equation}
    \left( E_{\pm}(\mathbf{k}) - \mu \right)  =  - \left( E_{\mp}(\mathbf{k}+ \mathbf{Q}) - \mu \right).
\end{equation}
Thereby, in addition to the Cooper instability in e.g. the singlet s-wave channel,  we observe an  antiferromagnetic instability in the particle-hole channel. Owing to the attractive retarded nature of the interaction one will expect charge-density-wave order to dominate\,\cite{Noack94}.

The $t=0$ point is special,  in the sense that the dispersion relation  in the vicinity of the chemical potential becomes that of Dirac fermions,
\begin{equation}
    E(\mathbf{k}) = \pm 2\alpha \sqrt{\sin^2(k_x)  + \sin^2(k_y)}.
\end{equation}
Expanding  around the  four time-reversal nodal points $\mathbf{K} = (0,0), (\pi,\pi), (0,\pi)$, and $(\pi,0)$ yields
\begin{equation}
\sin(k_x)\simeq s_x(\mathbf{K})\,p_x, \qquad
\sin(k_y)\simeq s_y(\mathbf{K})\,p_y,
\end{equation}
with node-dependent signs
\[
\begin{array}{c|cccc}
\mathbf{K} & (0,0) & (\pi,\pi) & (0,\pi) & (\pi,0)\\\hline
s_x & +1 & -1 & +1 & -1\\
s_y & +1 & -1 & -1 & +1
\end{array}
\]
Therefore, it leads to the following Dirac-like Hamiltonian 
\begin{equation}
\mathcal{H}_{0,\alpha} \simeq
v_F \sum_{\mathbf{p}} c^\dagger_{\mathbf{p}}
\Big[s_x(\mathbf{K})\,p_x\,\sigma^y - s_y(\mathbf{K})\,p_y\,\sigma^x\Big]\,
c^{\phantom\dagger}_{\mathbf{p}},
\end{equation}
where $\mathbf{p}=(p_x,p_y)$ measures momentum around the nodal points and the Fermi velocity is given by $v_F = 2 \alpha$.

We may relabel the nodes as $\mathbf{K}=(0,0) \leftrightarrow \tau=1, \mu =1 $;  $\mathbf{K}=(0,\pi) \leftrightarrow \tau=1, \mu =-1$; 
$\mathbf{K}=(\pi,\pi) \leftrightarrow \tau=-1, \mu =1 $;  and $\mathbf{K}=(\pi,0) \leftrightarrow \tau=-1, \mu =-1 $.
This allow us to introduce Pauli matrices $\tau^z$ and $\mu^z$, so that
\begin{equation}
s_x(\mathbf{K})=\tau^z_{\tau,\tau^{\prime}}, \qquad s_y(\mathbf{K})=\tau^z_{\tau,\tau^{\prime}}\mu^z_{\mu,\mu^{\prime}} ,
\end{equation}
which, in turn, leads to the compact Dirac-like Hamiltonian
\begin{eqnarray}
    {\cal H }_{0,\alpha}  & & =    v_F \sum_{\mathbf{p}} c^{\dagger}_{\mathbf{p},\tau,\mu,\sigma}\left( 
        p_x \delta_{\mu,\mu'}\tau^{z}_{\tau,\tau'}\sigma^{y}_{\sigma,\sigma'}  \right. 
         \nonumber \\
        & & \left.  - p_y \mu^{z}_{\mu,\mu'}\tau^{z}_{\tau,\tau'}\sigma^{x}_{\sigma,\sigma'}
    \right) c^{\phantom\dagger}_{\mathbf{p},\tau',\mu',\sigma'}  \nonumber \\ 
    & & \equiv 
    v_F \sum_{\mathbf{p}} c^{\dagger}_{\mathbf{p}} \left( p_x \tau^z \sigma^y - p_y \mu^z \tau^z \sigma^x \right)c^{\phantom\dagger}_{\mathbf{p}} ~.
\end{eqnarray}
In the last line we have suppressed the  indices  to  lighten the notation.
While on  the lattice, spin-orbit symmetry is discrete, it  becomes continuous  in the continuum limit  with generators  $(\tau^x \sigma^z, \tau^y \sigma^z,  \tau^z) $  of $SU(2)$.

Notice that, any finite hopping $t$ shifts the nodal points at $\mathbf{K}=(0,0)$ and $(\pi,\pi)$ away from the half-filling. In the continuum approximation, this hopping term acts as Dirac cone dependent chemical potential. It can be showed by expanding the hopping term around the nodal points, which leads to
\begin{equation}
    {\cal H }_{t,0}  =   - 4 t \sum_{\mathbf{p}} c^{\dagger}_{\mathbf{p}} \tau^z \left(\frac{\mu^z +1}{2}\right)c^{\phantom\dagger}_{\mathbf{p}}.
\end{equation}

To account for  superconducting as well as  charge-density-wave mass terms,  we adopt a Bogoliubov basis
\begin{equation}
        \eta^{\dagger}_{\mathbf{p}, \mu,\tau,\sigma, \nu } =  
        \left\{
        \begin{matrix}
            c^{\dagger}_{\phantom{-}\mathbf{p}, \mu,\tau,\sigma} &  \nu =\phantom{-} 1 \\
            c^{\phantom\dagger}_{-\mathbf{p}, \mu,\tau,\sigma}  & \nu = -1
        \end{matrix}
        \right.
\end{equation}
which, leads to
\begin{equation}\label{Eq:BdG_Hamil}
    {\cal H }_{0,\alpha} = v_F \sum_{\mathbf{p}} \eta^{\dagger}_{\mathbf{p} } \nu^z \left( 
        p_x \tau^z \sigma^y  - p_y \tau^z \mu^z \sigma^x
     \right)
    \eta^{\phantom\dagger}_{\mathbf{p} }. 
\end{equation}
The $s$-wave SC order parameter $\Delta$ and the 
CDW one, $\Phi$, form a triplet of anticommuting mass terms 
\begin{equation}
        \mathbf{M} = (\text{Im} \Delta, \text{Re} \Delta, \Phi  ) =  \left( \nu^x \sigma^y, \nu^y \sigma^y, \nu^z \tau^x  \right). 
\end{equation}  
That is, these three masses mutually anticommute, and also anticommutes with the Dirac-like Hamiltonian in Eq.\,\eqref{Eq:BdG_Hamil}, which eventually produces a gapped dispersion.
By defining the generators
\begin{equation}
  \Gamma_{nm} \;\equiv\; \frac{i}{2}\,[M_n,M_m], \qquad n,m = \{1,2,3\},
\end{equation}
one finds
$\Gamma_{nm} = \varepsilon_{nmk}\, M_k$,
which satisfy the $SO(3)$ algebra and act on the triplet as vector rotations
$[\Gamma_{ij}, M_k]   = i\,\varepsilon_{ijk}\, M_i$.
Hence, the pure Rashba term does not favour SC over CDW (or vice-versa), and provided that the interaction possesses the same symmetry then we expect degenerate SC and CDW fluctuations.

At half-filling the attractive Hubbard interaction possesses an $\mathrm{SO}(4)\!\cong\!\mathrm{SU}(2)_{\text{spin}}\!\times\!\mathrm{SU}(2)_{\eta}$ symmetry; the latter rotates the SC state onto the CDW one\,\cite{Yang90}. The spin symmetry is clearly broken by the Rashba spin-orbit coupling,  but  at $t=0$ SU(2)$_\eta$  symmetry is present both in the interaction term as well as is the Dirac-like  Hamiltonian.  Hence, as the appropriate Hamiltonian reads 
\begin{equation}
    {\cal H} = {\cal H}_{0,\alpha}  - U  \int_V d^2 \mathbf{x} \sum_{i=1}^{3} \left( c^{\dagger}_{\mathbf{x}} M_i c^{\phantom\dagger}_{\mathbf{x}} \right)^2 ~,
\end{equation} 
then it has SO(3) symmetry that renders SC and CDW states degenerate for any $U$ value. On the other hand,  at finite phonon frequencies, the instantaneous Hubbard term gives way to a retarded interaction.  This breaks the SO(3) symmetry and may favor CDW order\,\cite{Noack94}.

\subsection{The method}
We analyze the Hamiltonian in Eq.\,\eqref{Eq1} by employing a finite-temperature determinant Quantum Monte Carlo (DQMC) methodology \cite{Blankenbecler1981,Hirsch1985,rrds2003,Assaad2008,Gubernatis16}. In this approach, noncommuting terms of the Hamiltonian in the partition function are decoupled through the Trotter-Suzuki decomposition. It leads to the introduction of an imaginary-time axis by the discretization of the inverse temperature, \(\beta = L_\tau \Delta \tau\), with $L_\tau$ and \(\Delta \tau \) being the number of time slices and the discretization step, respectively. Notice that the Trotter-Suzuki decomposition introduces an error proportional to \((\Delta\tau)^2\); here, we define \(\Delta \tau =0.1 \), which is small enough ensure that systematic errors are smaller than statistical ones.

The Hamiltonian can be written as \( \mathcal{H} = \mathcal{H}_{k} + \mathcal{H}_{ph} + \mathcal{H}_{el-ph}  \), with terms on the right-hand side of the equality corresponding to the kinetic energy electrons (which includes the chemical potential and the RSOC contributions), the bare phonons energy, and the electron-phonon coupling, respectively. Therefore, the Trotter-Suzuki decomposition leads to the partition function
\[
Z = \text{Tr} \left[ \cdots e^{-\Delta\tau \mathcal{H}_{k}} e^{-\Delta\tau \mathcal{H}_{ph}} e^{-\Delta\tau \mathcal{H}_{el-ph}}  \cdots \right],
\]
where the trace is performed over both bosonic and fermionic degrees of freedom.
Using the HO's positions and the creation and annihilation operators as basis for the states, the bosonic trace leads to the inclusion of a bare phonons action, while the fermionic one leads to a single determinant. Notice that, due to the spin-flip terms of the RSOC, the fermionic weights can not be rewritten as a product of up and down determinants~\cite{Hirsch1985,White1988}.
That is, the partition function becomes
\[
Z = \int {\rm d} \{ x_{\mathbf{i},l} \} e^{-\Delta\tau S_{B}}  \left[ \text{det} \left( I + B_{L_{\tau}} B_{L_{\tau}-1} \cdots \right) \right]~,
\]
with $$ B_{l} = e^{[-\Delta\tau H_{k}]} e^{[-\Delta\tau H_{el-ph}(\{x_{\mathbf{i},l}\})]}  $$ being the products of the exponential of the matricial representation of the kinetic and electron-phonon terms, respectively, at a given time slice $l$. Here, $\{ x_{i,l} \}$ denotes the set of auxiliary (phonon) fields in real and imaginary-time coordinates, with the integral (i.e., the bosonic trace) \(\int d\{ x_{\mathbf{i},l} \}\)  being performed by Monte Carlo methods.
Finally, the exponential \( e^{-\Delta \tau S_{B}} \) is the bare phonon action
\[
S_{B} = \frac{\omega_{0}^{2}}{2} \sum_{i} \sum_{l=1}^{L_{\tau}} \left[ \frac{1}{\omega_{0}^{2} \Delta \tau^{2}} \left( x_{i,l} - x_{i,l+1} \right)^{2} + x^{2}_{i,l} \right],
\]
obtained from free phonon contribution in the partition function, i.e.~$e^{-\Delta\tau \mathcal{H}_{ph}}$. Therefore, $e^{-\Delta\tau S_{B}}  \text{det} \left( I + B_{L_{\tau}} B_{L_{\tau}-1} \cdots \right) $ is used as our statistical weight throughout the Monte Carlo sampling. This approach does not suffer with the infamous minus-sign problem \cite{Hirsch1985,White1988}.

\begin{figure}[t]
    \centering
    \includegraphics[scale=0.48]{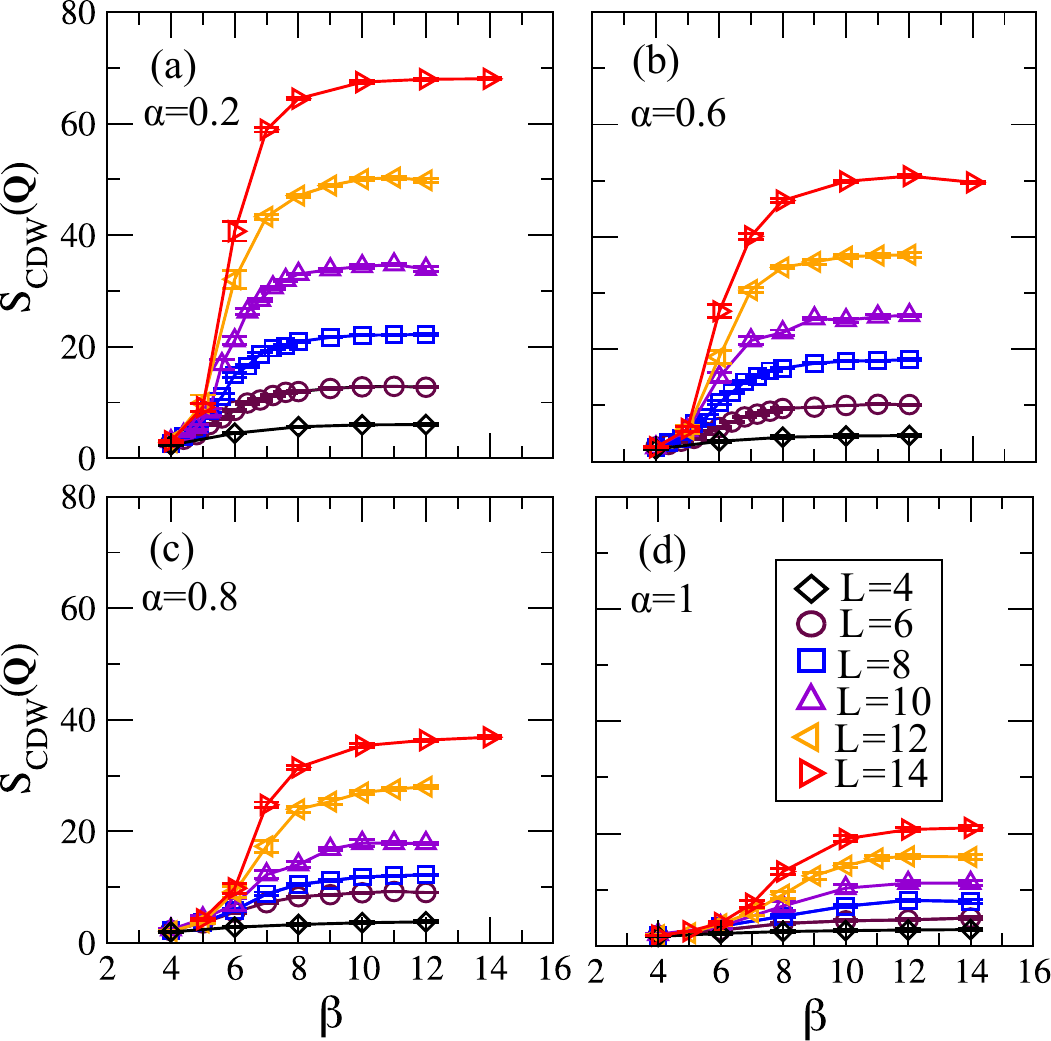}  
    \caption{The change-density wave structure factor, $S_{CDW}$(\textbf{Q}), as a function of the inverse of temperature ($\beta$) for different values system sizes and fixed (a) $\alpha=0.2$, (b) $\alpha=0.6$, (c) $\alpha=0.8$ and (d) $\alpha=1.0$. The phonon frequency is $\omega_{0}/t=1$ and the electron-phonon coupling is given by $\lambda/t=2$. Here, and in all subsequent figures, when not shown, error bars are smaller than symbol size.}
    \label{SvsB}
\end{figure}

To examine the effects of the electron-phonon coupling and the possible emergence of long-range order, we investigate charge-charge correlations, $\langle n_{\mathbf{i}} n_{\mathbf{j}} \rangle$, and their Fourier transform, the charge structure factor, 
\begin{equation}
    S_{\text{CDW}}(\textbf{q}) = \frac{1}{L^2} \sum_{\mathbf{i},\mathbf{j}} e^{i (\mathbf{i} - \mathbf{j}) \cdot \textbf{q}} \langle n_{\mathbf{i}} n_{\mathbf{j}} \rangle,
    \label{Eq2}
\end{equation}
with $L$ being the linear size of the system. The peak of \( S_{\text{CDW}}(\textbf{q}) \) defines the leading wavevector \(\textbf{Q}\) in which the system may exhibit charge order. At this point, we should warn that, to compute the correlation functions, one has to include spin-flip terms in the Wick's decomposition. For instance,
\begin{align}\label{Eq:Wicks}
\langle  n_{\mathbf{i}, \sigma} n_{\mathbf{j}, \bar{\sigma}} \rangle & = \langle c^{\dagger}_{\mathbf{i},\sigma} c_{\mathbf{i},\sigma} c^{\dagger}_{\mathbf{j},\bar{\sigma}} c_{\mathbf{j},\bar{\sigma}} \rangle \nonumber \\
& = \langle c^{\dagger}_{\mathbf{i},\sigma} c_{\mathbf{i},\sigma} \rangle \langle c^{\dagger}_{\mathbf{j},\bar{\sigma}} c_{\mathbf{j},\bar{\sigma}} \rangle + \langle c^{\dagger}_{\mathbf{i},\sigma} c_{\mathbf{j},\bar{\sigma}} \rangle \langle c_{\mathbf{i},\sigma} c^{\dagger}_{\mathbf{j},\bar{\sigma}} \rangle \nonumber \\
& = (1 - G^{\sigma,\sigma}_{\mathbf{i},\mathbf{i}})(1 - G^{\bar{\sigma},\bar{\sigma}}_{\mathbf{j},\mathbf{j}}) - G^{\sigma,\bar{\sigma}}_{\mathbf{i},\mathbf{j}}G^{\bar{\sigma},\sigma}_{\mathbf{j},\mathbf{i}}~,
\end{align}
in which $G^{\sigma,\bar{\sigma}}_{\mathbf{i},\mathbf{j}} \equiv \langle c_{\mathbf{i},\sigma} c^{\dagger}_{\mathbf{j},\bar{\sigma}} \rangle$ is the equal-time Green's function, with $\bar{\sigma} = -\sigma$.  

Given this, we determine the quantum critical points, as well as the finite temperature ones by the charge correlation ratio, defined as
\begin{equation}
    R_{CDW} = 1 - \frac{S_{\text{CDW}}(\textbf{Q} - \delta \textbf{q})}{S_{\text{CDW}}(\textbf{Q})}
    \label{Rcdw}
\end{equation}
where \(\textbf{Q} = (\pi,\pi)\) and \(\delta \textbf{q} = \frac{2\pi}{L}\). This quantity is related to the inverse of the width of the structure factor, which, in turn, is proportional to the correlation length. Therefore, in the thermodynamic limit, $R_{c} \rightarrow 1$ when long-range CDW occurs, while $R_{c} \rightarrow 0$ in the absence of it. As $R_c$ is a renormalization-group invariant, its crossing for different system sizes provides the critical points\,\cite{Kaul2015,Sato2018,Liu2018}.

\begin{figure}[t]
    \centering
    \includegraphics[scale=0.45]{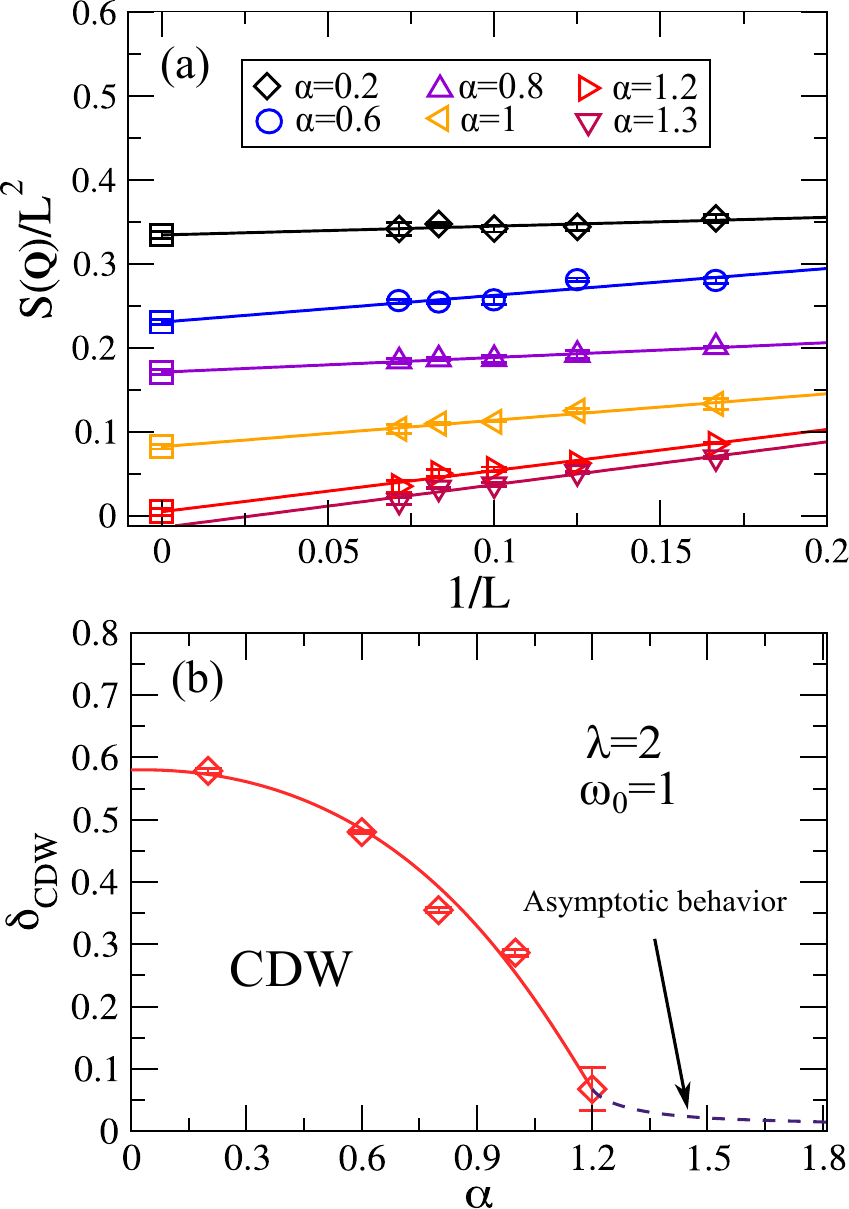}  \caption{(a) Finite-size scaling of the ground-state staggered CDW structure factor for different RSOC, $\alpha$, and fixed $\omega_{0}=1$ and $\lambda/t=2$. (b) The CDW order parameter, $\delta_{CDW}$, as a function of $\alpha$, for the same parameters of the previous panel. The red curve is just a guide to the eye.}
    \label{SL2vs1L}
\end{figure}

\section{Results and discussion}\label{results}

We begin our analysis by considering the case of fixed $\omega_{0}/t=1$ and EPC $\lambda/t = 2$, while varying the RSOC parameter, $\alpha$. We recall that the pure Holstein model has been extensively examined for this set of parameters, showing CDW order below a critical temperature $T_c/t = 0.174(2)$\,\cite{Batrouni19}. 
The leading effects of the Rashba coupling on this charge order are illustrated in Fig.\,\ref{SvsB}, where we show the behavior of the staggered CDW structure factor as a function of the inverse temperature ($\beta = 1/T$), for different values of $\alpha$ and system sizes. As the temperature decreases (i.e., as $\beta$ increases), the behavior of $S_{\text{CDW}}(\textbf{Q})$ changes from being weakly dependent on the system size at high temperatures (low $\beta$) to strongly dependent at low temperatures (high $\beta$). This transition occurs approximately at the critical $T_c (\alpha)$. However, as we are dealing with finite system sizes, at very low temperatures, the correlation length becomes larger than $L$, resulting in a plateaus in $S_{\text{CDW}}(\textbf{Q})$, as shown in Fig.\,\ref{SvsB}. Therefore, the reduction of the strength of the $S_{\text{CDW}}(\textbf{Q})$ plateuaus as $\alpha$ increases provides a clear evidence that Rashba coupling weakens the CDW phase. Moreover, as discussed in more details below, the energy scale at which $S_{\text{CDW}}(\textbf{Q})$ begins to increase is shifted to larger values of $\beta$, suggesting that the Rashba coupling not only affects the order parameter but also pushes the critical temperature to lower values.

To obtain the CDW order parameter, $\delta_{\text{CDW}}$, we assume that the low-$T$ plateaus of $S_{\text{CDW}}(\textbf{Q})$ provide a good estimate of the ground state response. Therefore, we extrapolate $\frac{S_{\rm CDW}(\textbf{Q})}{L^2}$ as a function of $1/L$ to capture its behavior in the thermodynamic limit. That is, we employ
\begin{equation}
   \frac{S_{\rm CDW}(\textbf{Q})}{L^2} =  \delta^{2}_{\text{CDW}} + \frac{A}{L} + \mathcal{O}(1/L^2)~,
    \label{OP1}
\end{equation}
where $A$ is a constant obtained from fitting the QMC data. This finite-size scaling analysis is shown in Fig.\,\ref{SL2vs1L}\,(a) for different values of RSOC. The error bar for $\delta^{2}_{\text{CDW}}$ comes from the linear fitting.

\begin{figure}[t]
\centering
\includegraphics[scale=0.42]{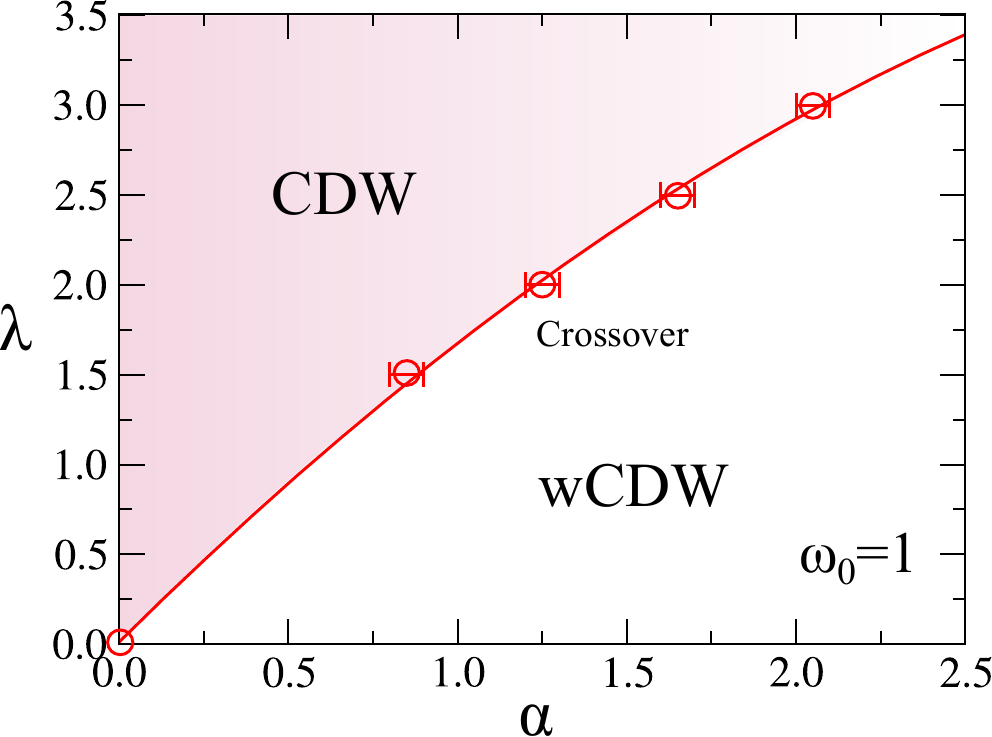}    \caption{The ground state phase diagram of the Rashba-Holstein model for fixed $\omega_{0}=1$. The solid red line is just a \textit{crossover} between a robust CDW (magenta region) and a weak CDW (white region) one. }
    \label{Fig6}
\end{figure}

In line with the previous discussion, Fig.\,\ref{SL2vs1L}\,(a) demonstrates that, in the thermodynamic limit, $S_{\rm CDW}(\textbf{Q})/L^2$ is significantly impacted by the RSOC, leading to nearly vanishing values for $\alpha/t \approx 1.3$ . These results allow us to present the CDW order parameter as a function of $\alpha/t$, as shown in Fig.\,\ref{SL2vs1L}\,(b). Notice that, for small values of RSOC, the CDW response remains similar to that of the pure Holstein model, but, as $\alpha$ increases, the order parameter is suppressed.
At this point, a remark should be made: although Fig.\,\ref{SL2vs1L} suggests a zero order parameter at $\alpha \approx 1.3$, we expect an asymptotic approach to zero instead, due to the instability of the particle-hole channel from the nesting of the Fermi surface.
Indeed, from a mean-field perspective, the CDW order parameter exhibits asymptotic behavior as $\alpha \to 0$ \cite{Fontenele2024}, analogous to the mean-field behavior of the antiferromagnetic order parameter in the repulsive Hubbard-Rashba model \cite{Kubo2024,Kawano2023}; in both cases, the ordering occurs at the wavevector $\mathbf{Q}=(\pi,\pi)$. However, in contrast to the repulsive Hubbard-Rashba model -- where increasing RSOC can stabilize additional magnetic orders --, our QMC data for the Holstein model do not indicate distinct CDW orderings over the parameter ranges explored.
That is, while the RSOC strongly affects spin ordering in the Hubbard model (in particular at strong coupling by means of a Dzyaloshinskii-Moriya interaction\,\cite{Cocks12,Radic12}), its interplay with lattice degrees of freedom in the Holstein model is insufficient to stabilize alternative CDW patterns.

\begin{figure}[t]
    \centering    \includegraphics[scale=0.33]{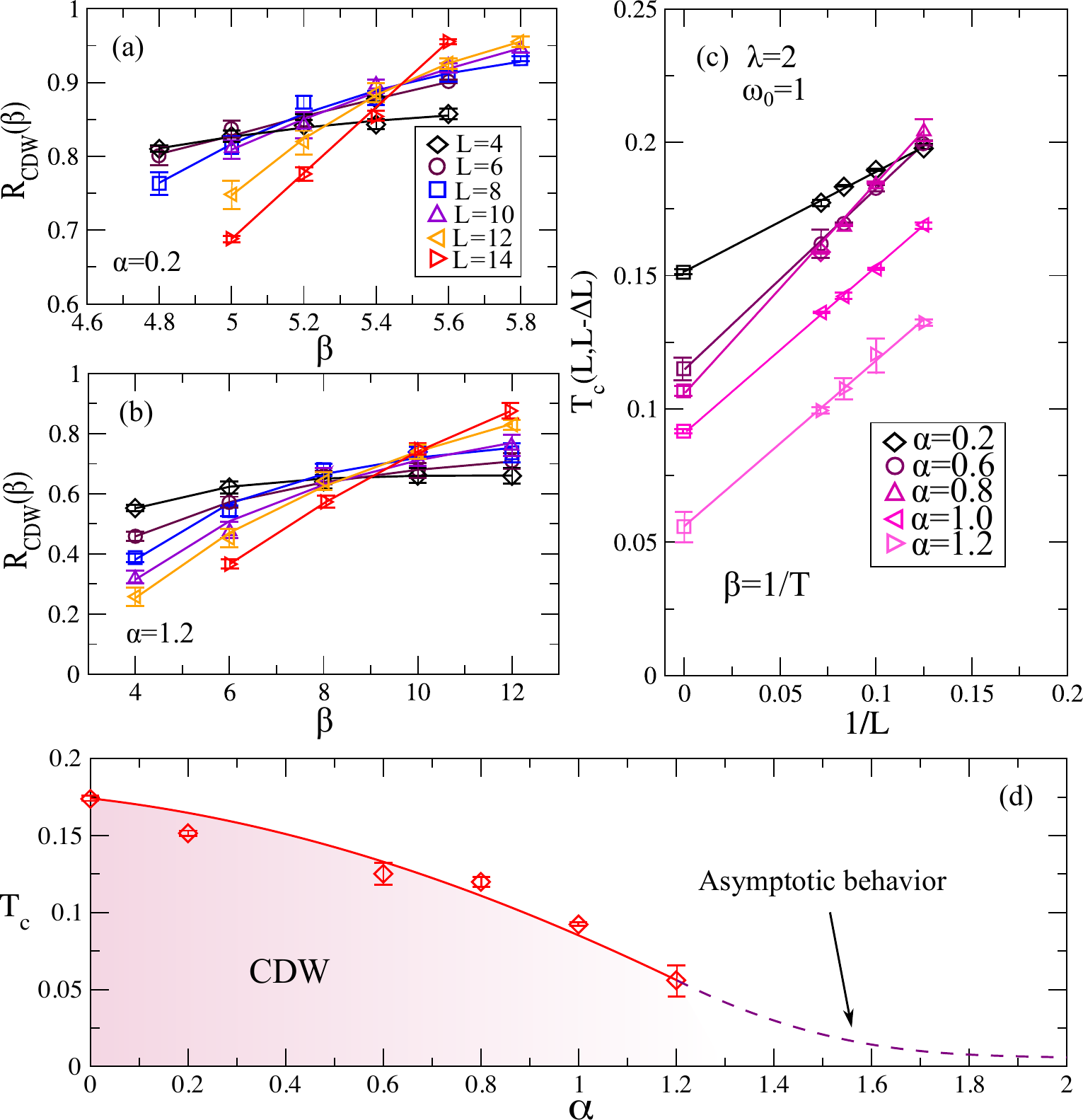}
    \caption{The CDW correlation ratio $R_{CDW}(\beta)$ as a function of the inverse temperature ($\beta$=1/T), for fixed (a) $\alpha= 0.2$ and (b) $\alpha= 1.2$, and different lattice sizes. (c) The extrapolation of the crossings of the $R_{CDW}(\beta)$, between $L$ and $L-\Delta L$ system sizes, $T_{c}(L, L - \Delta L)$. Here, we fixed $\Delta L =4$. (d) The extrapolated critical temperatures of the Rashba-Holstein model for fixed $\omega_0 /t =1$. The red curve is just a guide to the eye.}
    \label{RcBc}
\end{figure}

Then, the ground state of the Rashba-Holstein model is expected to always manifest as a CDW, for any $\alpha \geq 0$. However, the strength of its order parameter (which is proportional to the charge gap at the Fermi level) depends on the RSOC, becoming exponentially small for large values of $\alpha$. 
Here, we recall that the particle–hole and particle-particle channels exhibit similar logarithmic divergences. Therefore, in the regime of large RSOC, where we expect an exponentially small CDW order parameter, a competition between CDW and SC orders may occur, although we are unable to quantify it by QMC.
Thus, for practical purposes, it is useful to identify the crossover region where the system goes from a strong CDW regime to a weak CDW (wCDW) one, which can be marked at the point where the extrapolation of $S_{\rm CDW}(\textbf{Q})/L^2$ vanishes. Identifying this crossover is important, particularly because the region with such an exponentially small order parameter is unstable to even small perturbations. For instance, thermal fluctuations or further neighbor hopping terms (that break the nesting of the Fermi surface) can lead to the occurrence of a plain Rashba metal;
similarly, a small degree of disorder can lead to SC ordering.

\begin{figure}[t]
    \centering
    \includegraphics[scale=0.54]{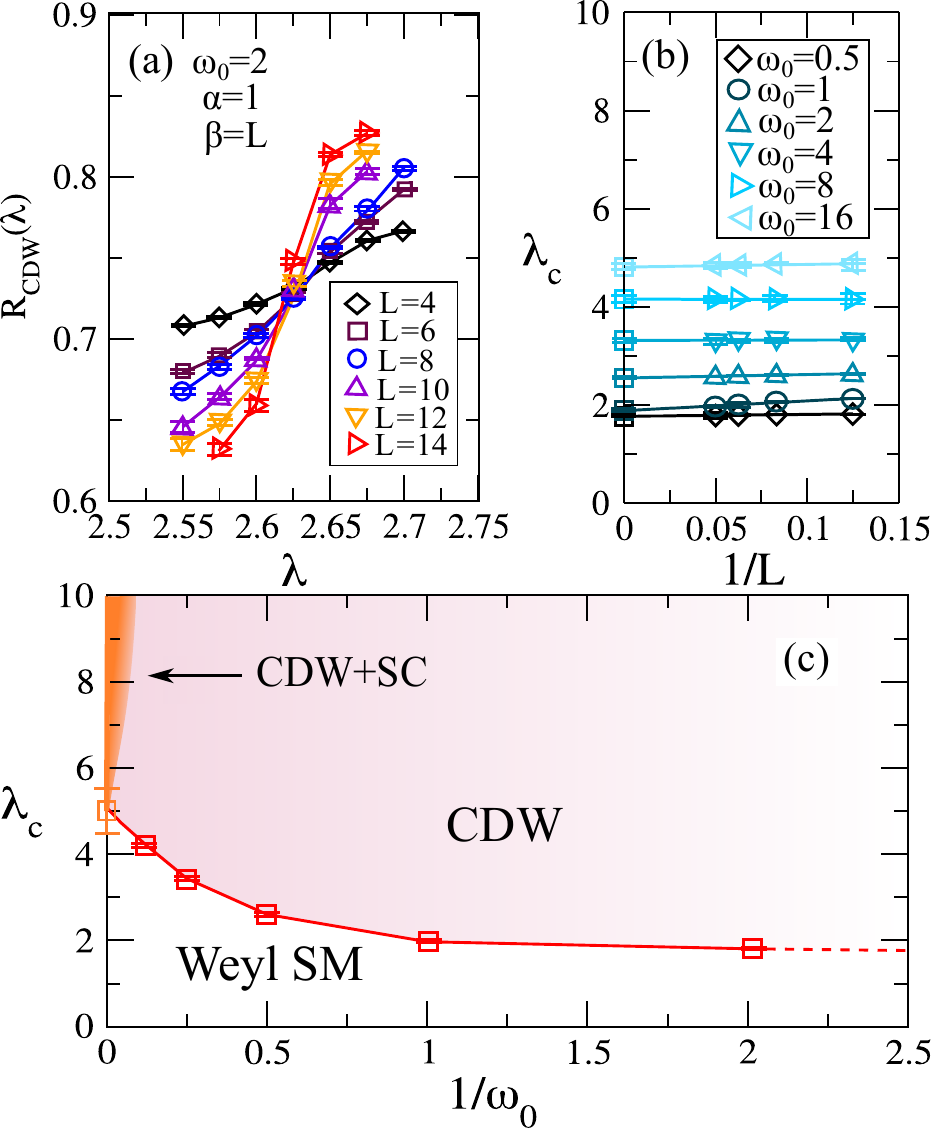}
    \caption{(a) The CDW correlation ratio, $R_{CDW}(\lambda)$, as a function of $\lambda$ in the limit of the pure Rashba hopping, and fixed $\omega_{0}/\alpha=2$, $\beta=L$, and $\alpha=1$. (b) The extrapolation of the crossings of the $R_{CDW}(\beta)$, between $L$ and $L+\Delta L$ system sizes. Here, we fixed $\Delta L =4$. (c) The phase diagram of the Rashba-Holstein model in the limit of a pure Rashba hopping. The coexistence between CDW and SC occurs for large frequencies, and become degenerate in the limit of $\omega_0 \to \infty$.}
    \label{Fig7}
\end{figure}

By repeating the same procedure outlined in Fig.\,\ref{SL2vs1L}, but for other values of EPC, one obtains the phase diagram of Fig.\,\ref{Fig6}. Here, as already mentioned, the QMC points corresponds to the vanishing $\delta_{\rm CDW}$ in the thermodynamic limit, and determines the CDW-wCDW crossover. However, for a fixed low temperature (e.g., $\beta/t=40$), this crossover line becomes a second order transition belonging to the two-dimensional Ising universality class, from a CDW phase to a Rashba metal.

\begin{figure}[t]
    \centering
   \includegraphics[scale=0.39]{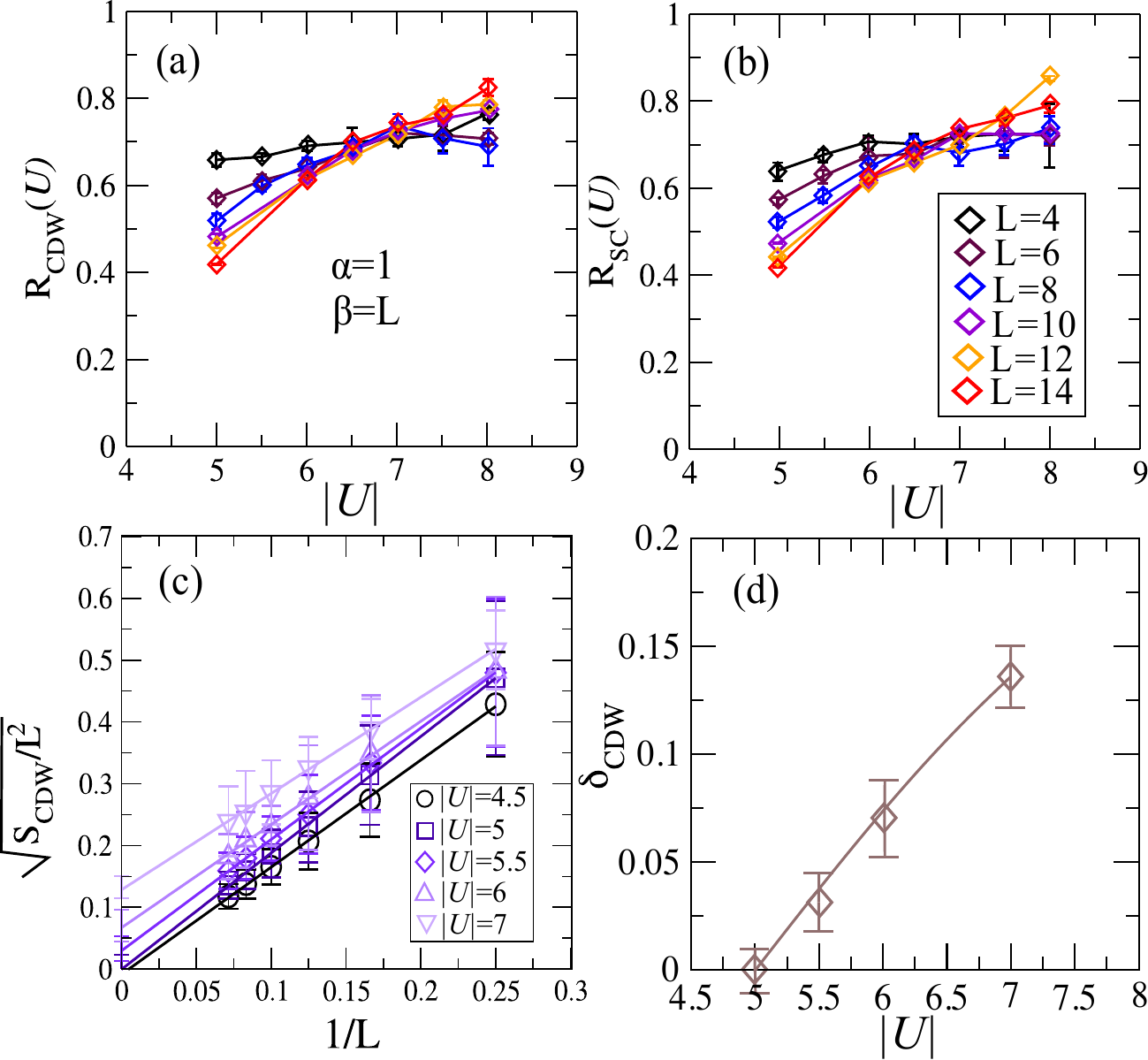}
    \caption{The (a) CDW and (b) SC correlation ratios as functions of $|U|$, in the limit of the attractive Hubbard model, for the case of a pure Rashba hopping. (c) The extrapolations to the thermodynamic limit of the $\sqrt{S_{CDW}/\text{L}^{2}}$, which leads to the order parameter presented in panel (d). The critical point is at $|U_{c}|=5.0 \pm 0.5$. }
    \label{Fig8}
\end{figure}

It is also important to examine the finite temperature behavior, particularly to identify the critical temperature for the emergence of CDW, which is determined by the crossings of the correlation ratio, Eq.\,\eqref{Rcdw}. Figure \ref{RcBc} shows the behavior of $R_{\text{CDW}}$ as a function of $\beta$ for different lattice sizes, and fixed (a) $\alpha/t=0.2$ and (b) $\alpha/t=1.2$. In all cases, the behavior is similar: $R_{\text{CDW}} \to 0$ as $\beta \to 0$, and $R_{\text{CDW}} \to 1$ as $\beta \to \infty$. However, it is the crossing points for different system sizes that determine the critical value $\beta_c$. For $\alpha/t=0.2$, these crossings occur around $\beta \approx 5.4$, while for $\alpha/t=1.2$, they occur at larger values of $\beta$, as expected. A more precise estimate of $T_c$ can be obtained by extrapolating these crossings to the thermodynamic limit. This is done by determining the crossing points between $R_{\text{CDW}}(L)$ and $R_{\text{CDW}}(L-\Delta L)$, with $\Delta L = 4$ -- defining this as $T_{c}(L,L -\Delta L)$ -- and then extrapolating to $1/L \rightarrow 0$, as illustrated in Fig.\,\ref{RcBc}\,(c).
The results for $T_c$ in the thermodynamic limit are presented in Fig.\,\ref{RcBc}\,(d), where one may notice that the larger $\alpha$, the lower $T_c$. An asymptotic behavior is also expected for $ T_c $; however, for $\alpha/t \gtrsim 1.2$, achieving this would require $\beta t \gg 20$, which is computationally costly.

A similar asymptotic behavior is expected to occur for a different set of parameters, as long as instabilities in the particle-hole channel persist. However, this situation changes in the limit of the pure Rashba model, i.e., when $t/\alpha = 0$. As discussed earlier, in this limit there are four Weyl cones at $(0,0)$, $(\pi,0)$, $(0,\pi)$, and $(\pi,\pi)$ for the non-interacting half-filled case. Similar to the honeycomb lattice, this vanishing density of states requires sufficiently strong electronic interactions to lead to a quantum phase transition\,\cite{Paiva05,Sorella12,Assaad13}. Indeed, a semimetal-CDW quantum critical point (QCP) occurs for the Holstein and Hubbard-Holstein models in the half-filled honeycomb lattice\,\cite{Zhang20192,Chen2019,Costa2021}, thus we expect an analogous behavior in  this limit of the pure Rashba coupling. 

We proceed by projecting onto the ground state, assuming Lorentz invariance and defining $\beta \propto L^z$, where $z = 1$ is the dynamic critical exponent at the QCP \,\cite{Assaad13,Otsuka16,Toldin15}. As before, the critical point is determined by the crossing of $R_{\text{CDW}}$ for different lattice sizes, as shown in Fig.\,\ref{Fig7}\,(a). In this case, we have fixed $\beta = L$ and $\omega_{0} / \alpha = 2$, while varying $\lambda$, with clear crossings around $\lambda / \alpha \approx 2.62$. Following a similar procedure as that used for determining the critical temperatures, we extrapolate the crossings for the critical $\lambda_c$ to the thermodynamic limit, as presented in Fig.\,\ref{Fig7}\,(b), thereby determining the QCP. We repeated this analysis for other values of $\omega_{0}$ [see, e.g., Fig.\,\ref{Fig7}\,(b)], which allowed us to construct the ground state phase diagram of the Rashba-Holstein model in the limit of the Weyl semimetal cones (i.e.~a pure Rashba hopping), depicted in Fig.\,\ref{Fig7}\,(c).
Notice that, even for high frequency (e.g., $\omega/\alpha = 16$) the ground state is either a staggered CDW or a WSM one, with no enhancement of SC or striped CDW phases.

In the adiabatic limit ($\omega_0 \to 0$), the Holstein model can be solved using a mean-field approach, as it corresponds to permanent (static) distortions in the lattice. In contrast, in the antiadiabatic limit ($\omega_0 \to \infty$), the indirect interaction between electrons becomes instantaneous, and by integrating out the phonons, one obtains the attractive Hubbard model with $|U| = \lambda$. Both limits have been previously investigated in the presence of RSOC, as discussed in Refs.\,\onlinecite{Tang2014,Rosenberg2017,Song2024, Fontenele2024}.
Despite this, the attractive Hubbard model with a pure Rashba hopping (i.e.~$t/ \alpha =0$) has never been explored in the literature. Therefore, to complete the phase diagram in Fig.\,\ref{Fig7}\,(c), it is necessary to examine the emergence of CDW and/or SC in the limit of $\omega_0 \to \infty$.

In view of this, we employ the same ground state projection (i.e., $\beta \propto L^z$) for the attractive Hubbard model with pure Rashba hopping. Figures \ref{Fig8}\,(a) and (b) show the results for $R_{\rm CDW}$ and $R_{\rm SC}$ as functions of $|U|$, respectively. Here, $R_{\rm SC}$ is defined similarly to Eq.\,\eqref{Rcdw}, but it refers to the pairing structure factor with $\mathbf{Q}=0$, instead of the staggered one.
Notice that both CDW and SC responses are identical within error bars.
As discussed in Sec.\,\ref{SubSection:Symmetries}, although the $SU(2)$ symmetry is broken in the presence of RSOC, in the limit of a pure Rashba hopping (i.e.~$t=0$), the $SU(2)-\eta$ symmetry is recovered in the attractive Hubbard model.
Therefore, the degeneracy presented in Figs.\,\ref{Fig8}\,(a) and (b) are imposed by symmetry.
Given this, we identify the QCP at $|U_c|\approx 5$, as shown in Fig.\,\ref{Fig8}\,(c), with the CDW (and SC) order parameters displayed in Fig.\,\ref{Fig8}\,(d).

\begin{figure}[t]
    \centering
    \includegraphics[scale=0.50]{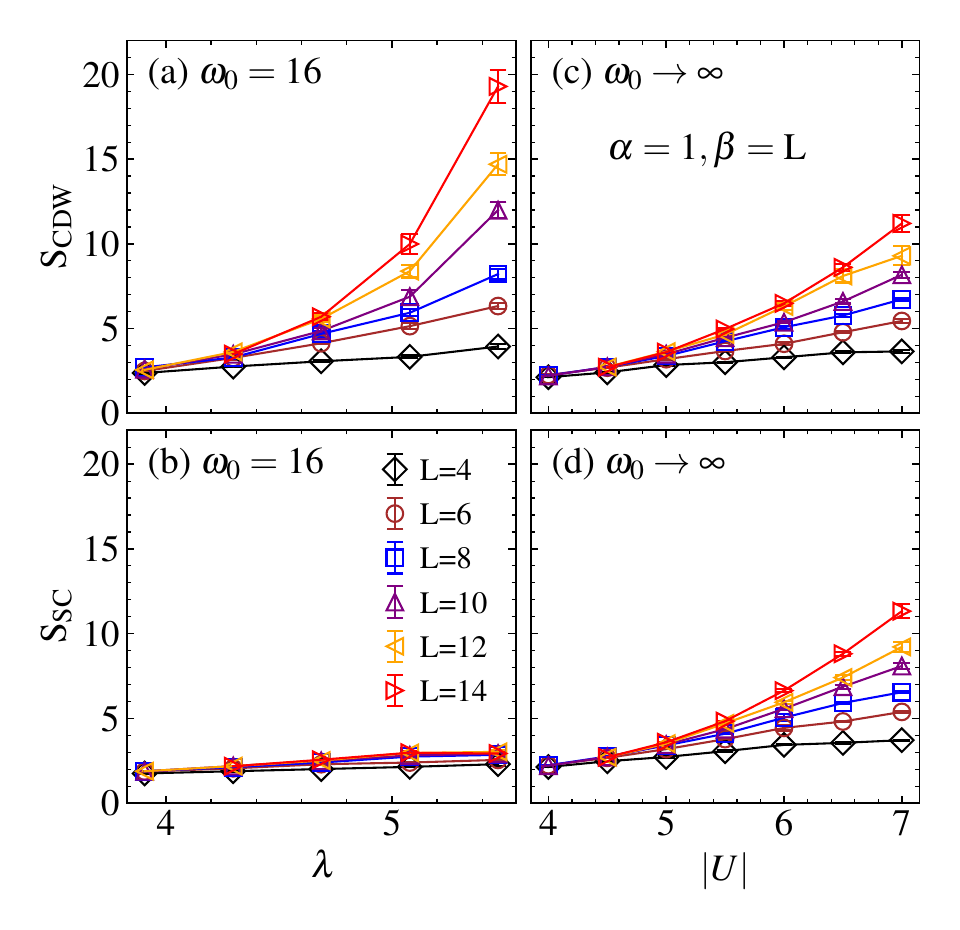}
    \caption{ (a) Staggered CDW structure factor and (b) homogeneous pair structure factor as functions of the electron-phonon coupling parameter, $\lambda$, for fixed $\omega_0 / \alpha = 16$ and different lattice sizes. Panels (c) and (d) present the corresponding quantities for the attractive Hubbard model, and as functions of $|U|$. Both cases are in the limit of a pure Rashba hopping.}
    \label{Fig:w16}
\end{figure}

Finally, the previous results naturally prompt the question of whether the coexistence between CDW and SC occurs only in the limit $\omega_0 \to \infty$ (symmetry-imposed), or if it can also manifest at finite frequency values (not symmetry-imposed). To address this issue, we fix $\omega_0/\alpha = 16$ and $\beta = L$, while varying $\lambda$. The results are displayed in Fig.\,\ref{Fig:w16}, for the response of (a) the staggered CDW structure factor and (b) the homogeneous $s$-wave SC structure factor, $S_{\rm SC}$. While the former shows significant enhancement for $\lambda \gtrsim 4.5$, consistent with a charge ordered ground state, the latter exhibits only weak size dependence as $\lambda$ increases, indicating the absence of SC long-range order. This should be compared to the case of the attractive Hubbard model, shown in Figs.\,\ref{Fig:w16}\,(c) and (d), for $S_{\rm CDW}$ and $S_{\rm SC}$, respectively. In stark contrast, both the CDW and SC structure factors exhibit significant finite-size dependence, suggesting their coexistence in this regime.

\begin{figure}[t]
    \centering
–    \includegraphics[scale=0.54]{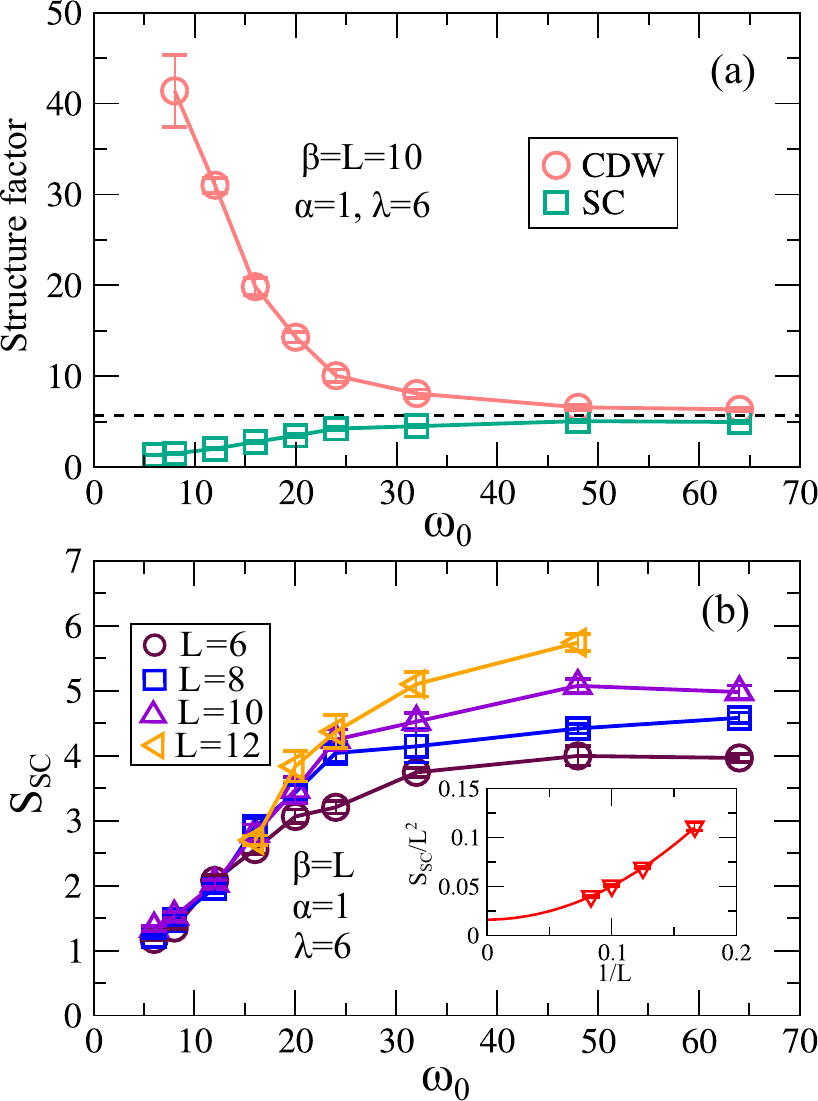}
    \caption{(a) Charge and pair structure factors as a function of the phonon frequency $\omega_{0}$ at fixed $\beta=L=10$ and $\lambda/\alpha=6$. The horizontal dashed line marks the corresponding degenerate $S_{\rm CDW}$ and $S_{\rm SC}$ value in the antiadiabatic limit $(\omega_{0} \to \infty)$ for the same parameters. (b) Pair structure factor as a function of $\omega_{0}$ for several lattice sizes $L$, with fixed $\beta=L$ and $\lambda/\alpha=6$. Inset: extrapolation of $S_{\rm sc}/L^2$ to the thermodynamic limit, for fixed $\omega_{0} = 48$. For both panels, we set $t =0$ and $\alpha =1$.}
    \label{Fig:antiadiabaticlim}
\end{figure}

The previous analysis suggests that any CDW-SC coexistence at finite $\omega_{0}$ would be restricted to large values of $\omega_{0}$, in the regime of pure Rashba hopping. To estimate the required scale, Fig.\,\ref{Fig:antiadiabaticlim}\,(a) compares $S_{\mathrm{CDW}}$ and $S_{\mathrm{SC}}$ as functions of $\omega_{0}$. Both quantities trend toward the attractive Hubbard limit (horizontal dashed line) and become degenerate with increasing frequency. However, notice that achieving (results quantitatively consistent with) the antiadiabatic limit requires $\omega_{0}/\alpha \gtrsim 64$, i.e.~$\omega_{0}$ roughly an order of magnitude larger than the bandwidth. With this in mind, we turn to examine the behavior of the SC order parameter at large frequencies. Figure \ref{Fig:antiadiabaticlim}\,(b) shows $S_{\mathrm{SC}}$ as a function of $\omega_{0}$ for several lattice sizes. Notice that, for $\omega_{0}/\alpha \gtrsim 24$, $S_{\mathrm{SC}}$ increases with system size, which is consistent with the occurrence of long-range order. Extrapolating these data for large frequencies to the thermodynamic limit leads a nonzero SC order parameter (as showed in inset). This confirms coexistence of SC and CDW order at finite $\omega_{0}$, while the two become degenerate (imposed by symmetry) in the limit of $\omega_{0}\to\infty$\,\footnote{We expect that a similar behavior should occur even for $t/\alpha \neq 0$, since both non-interacting particle-hole and particle-particle channels diverges logarithmically.}.

\section{Conclusions}\label{conclusion}

In this work, we investigate the role of RSOC in the emergence of CDW or SC phases in the Holstein model on a half-filled square lattice. By seeking to go beyond mean-field approaches, we employed unbiased finite-temperature Quantum Monte Carlo simulations. Within this methodology, we determined the behavior of the ground state order parameter as a function of RSOC [as shown in Fig.\,\ref{SL2vs1L}\,(c)], which allowed us to construct the phase diagram in Fig.\,\ref{Fig6}. Our results suggest that, due to instabilities in the particle-hole channel, the Rashba metal is unstable, and the emergence of a CDW phase is expected for any RSOC, although the CDW phase remains relatively weak when $\alpha \gg t$. Additionally, we present the finite-temperature phase diagram in Fig.\,\ref{RcBc}\,(d), where similar asymptotic behavior is also expected at large values of RSOC.

However, the features of the model change drastically in the limit of a pure Rashba hopping, i.e., when $t/\alpha = 0$. In this limit, four Weyl cones appear at the half-filling, and quantum phase transitions are expected to occur at strong interactions. Specifically, we demonstrated that the CDW phase emerges for a finite quantum critical coupling $\lambda_c$, for a given phonon frequency. By varying $\omega_0$, the critical $\lambda_c$ also changes, leading to the phase diagram shown in Fig.\,\ref{Fig7}.  
At finite phonon frequencies, we  expect this transition to belong  to the Ising Gross-Neveu universality class with  $N_f = 8$  two component Weyl cones\,\cite{Herbut09a,Huffman19}. 
In the antiadiabatic limit 
where  $\omega_0 \to \infty$ the interaction corresponds to the attractive Hubbard model, with $|U| = \lambda$. In this limit, and as described in Sec.~\ref{SubSection:Symmetries} we expect an emergent 
SU(2)$_{\eta} $   symmetry  the ties SC and CDW orders together.
For large frequencies ($\omega_{0} \gtrsim 24$), SC emerges, with the ground state exhibiting a coexistence between CDW and SC. These quantities become degenerate when $\omega_0 \to \infty$.
This corresponds to the same  universality class  as that of the Hubbard model on the honeycomb lattice \cite{Assaad13,Otsuka16}.
These findings provide a step forward in understanding the interplay between SC and CDW phases in systems with spin-orbit coupling, and we hope they could shed light on the nature of these compounds.

\color{black}

\section*{ACKNOWLEDGMENTS}
The authors are grateful to the Brazilian Agencies Conselho Nacional de Desenvolvimento Cient\'ifico e
Tecnol\'ogico (CNPq), Coordena\c c\~ao de Aperfei\c coamento de Pessoal de Ensino Superior (CAPES), and Fundação de Amparo \`a Pesquisa do Estado do Rio de Janeiro, FAPERJ.
J.F.~thanks FAPERJ, Grant No.~SEI-260003/019642/2022, and ANID Fondecyt grant No.\,3240320.
S.A.S.J.~thanks CNPq, grant No.~201000/2024-5.
N.C.C.~acknowledges support from FAPERJ Grants No.~E-26/200.258/2023 [SEI-260003/000623/2023] and E-26/210.592/2025 [SEI-260003/004500/2025], CNPq Grants No.~313065/2021-7 and 308130/2025-1, and Serrapilheira Institute Grant No.~R-2502-52037.
FA thanks the W\"urzburg-Dresden Cluster of Excellence on Complexity and Topology in Quantum Matter ct.qmat (EXC 2147, project-id 390858490).
This research was partially supported by the supercomputing infrastructure of the NLHPC (Grant No. CCSS210001).

\bibliography{ref}

\end{document}